\documentclass[twocolumn,showpacs,amsmath,amssymb]{revtex4}
\usepackage{graphicx}% Include figure files
\usepackage{dcolumn}% Align table columns on decimal point
\usepackage{bm}% bold math
\usepackage{psfig}% bold math
\newcommand{\ag}{a_{3g}}
\newcommand{\aga}{a_{\gamma}}
\newcommand{\ac}{a_c}
\newcommand{\BR}{{\cal B}}

\newcommand{\psp}{\psi^{\prime}}

\newcommand{\pspp}{\psi^{\prime \prime}}

\newcommand{\jpsi}{J/\psi}

\newcommand{\EE}{e^+e^-}

\newcommand{\PP}{\pi^+\pi^-}
\newcommand{\KK}{K^+K^-}
\newcommand{\pp}{\pi^+\pi^-}

\newcommand{\kskl}{K^0_SK^0_L}
\newcommand{\KKSC}{K^{*+}K^-}
\newcommand{\KKSN}{K^{*0}\overline{K^0}}

\newcommand{\OP}{\omega\pi^0}

\newcommand{\RET}{\rho\eta}
\newcommand{\RETp}{\rho\eta^{\prime}}
\newcommand{\OET}{\omega\eta}
\newcommand{\OETp}{\omega\eta^{\prime}}
\newcommand{\FET}{\phi\eta} 
\newcommand{\FETp}{\phi\eta^{\prime}}

\newcommand{\rhopi}{\rho\pi}

\newcommand{\ra}{\rightarrow}

\newcommand{\beq}{\begin{equation}}
\newcommand{\eeq}{\end{equation}}
\newcommand{\beqn}{\begin{eqnarray}}
\newcommand{\eeqn}{\end{eqnarray}}
\newcommand{\beqns}{\begin{eqnarray*}}
\newcommand{\eeqns}{\end{eqnarray*}}
\newcommand{\bfg}{\begin{figure}}
\newcommand{\efg}{\end{figure}}
\newcommand{\bitm}{\begin{itemize}}
\newcommand{\eitm}{\end{itemize}}
\newcommand{\bnum}{\begin{enumerate}}
\newcommand{\enum}{\end{enumerate}}
\newcommand{\btbl}{\begin{table}}
\newcommand{\etbl}{\end{table}}
\newcommand{\btbu}{\begin{tabular}}
\newcommand{\etbu}{\end{tabular}}
\def\eref#1{(\ref{#1})}
\def\Journal#1#2#3#4{{#1} {\bf #2}, #3 (#4)}
\def\NPB{Nucl. Phys. B}
\def\PLB{Phys. Lett. B}
\def\PRL{Phys. Rev. Lett.}
\def\PRD{Phys. Rev. D}
\def\PRP{Phys. Rep.}

\def\EPJC{Eur. Phys. J. C}
\def\JHEP{Journal of High Energy Physics}

\begin{document}

\preprint{Draft-PRD}

\title{Measurement of the exclusive light
hadron decays of the $\pspp$ in $\EE$ experiments}
\author{P. Wang}
\author{X.H. Mo}
\author{C.Z. Yuan}
\affiliation{Institute of High Energy Physics, Chinese Academy of
Sciences, Beijing 100049, China}

\date{\today}

\begin{abstract}

The measurement of the exclusive light hadron decays of 
the $\pspp$ in
$\EE$ experiments with significant interference between the
$\pspp$ and non-resonance continuum amplitudes is discussed. The
radiative correction and the Monte Carlo simulation are studied.
A possible scheme to verify the destructive interference is
proposed for the detectors with energy-momentum resolution of $(1
\sim 2)\%$. \\

{\em Keywords:} $\pspp$ exclusive decay; interference; 
Monte Carlo simulation; $\EE$ experiment.

\end{abstract}

\pacs{13.25.Gv, 12.38.Qk, 14.40.Gx}

\maketitle

\section{Introduction}

The study of the charmonium has been revived due to large data
samples collected by CLEOc and BESII as well as by the
$B$-factories. A prominent physics which has drawn interest for
more than two decades is the very small branching fractions of
$\rhopi$ and other vector-pseudoscalar modes in $\psp$ decays
compared with their large branching fractions in $\jpsi$ decays.
One proposal to solve this puzzle is the $2S$-$1D$ states mixing
scenario~\cite{rosner} which predicts enhanced rate of $\rhopi$
mode in $\pspp$ decays. This scenario is extended to other decay
modes besides $\rhopi$, it foresees possible large partial widthes
for the light hadron modes in $\pspp$ decays~\cite{wym7,wym1}.

Recently, CLEOc and BESII reported the search for various light
hadron decays of the $\pspp$~\cite{cleolp05,besvp,besrp}. These
experiments produce the $\pspp$ in $\EE$ collision so the radiative
correction must be taken into account in the data analysis.
Another feature of the data analysis of the resonances produced in
$\EE$ collision is the need to consider the non-resonance
continuum amplitude. For example, in the $\EE \ra \rhopi$ data
collected at the $\pspp$ mass, if the $2S$-$1D$ mixing scenario
gives correct ${\cal B}(\pspp \ra \rhopi)$, then the non-resonance
continuum amplitude is comparable to the $\pspp$ decay amplitude,
the measured cross section is the result of the interference of
the two~\cite{wym3}. For the narrow resonances like $\jpsi$,
$\psp$, $\Upsilon(1S)$, $\Upsilon(2S)$, and $\Upsilon(3S)$, the
energy spread of the $\EE$ colliders must also be considered. But
in this work, we concentrate on the wide resonances like $\pspp$,
of which the width is much wider than the finite energy spread of
the $\EE$ colliders, so the energy spread does not change the
observed cross section in any significant way. In this paper, we
pay special attention to the circumstance that the interference
between the amplitudes of the resonance and continuum has
substantial contribution, particularly the circumstance that such
interference is destructive. We present the characteristic
features of the radiative correction under such circumstance. In
the forthcoming sections, we begin with the amplitudes for the
observed processes in $\EE$ experiments and the parametrization of
them. Then we present the general properties of radiative
correction. Next we turn to the Monte Carlo simulation. We discuss
the invariant mass distribution of the final hadron systems for
the continuum process and for the pure resonance process, as well
as for the circumstance in which both the resonance and continuum
amplitudes exist and there is significant interference effect
between them. Finally, we propose a possible scheme to verify the
destructive interference with the data collected at the energy of
the $\pspp$ mass for a detector with the energy-momentum
resolution of $(1 \sim 2)\%$.

Avoiding complexity but without losing generality, we restrict our
discussions on two situations in which the data are collected
either off resonance at continuum, or at the energy of the
resonance mass. For the experimental setting in which $\pspp$ is
scanned, the technique details are more complicated. We shall
leave its study to a future work. Through out the paper, the
resonance $\pspp$ is taken as an example, but the analysis can be
easily extended to other resonances with their width to mass
ratios comparable to the energy-momentum resolution of the
detector.

\section{Three amplitudes in $\EE$ experiments}
\label{threeamplitude}

\begin{figure}
\begin{minipage}{8cm}
\includegraphics[width=3.25cm,height=2.5cm]{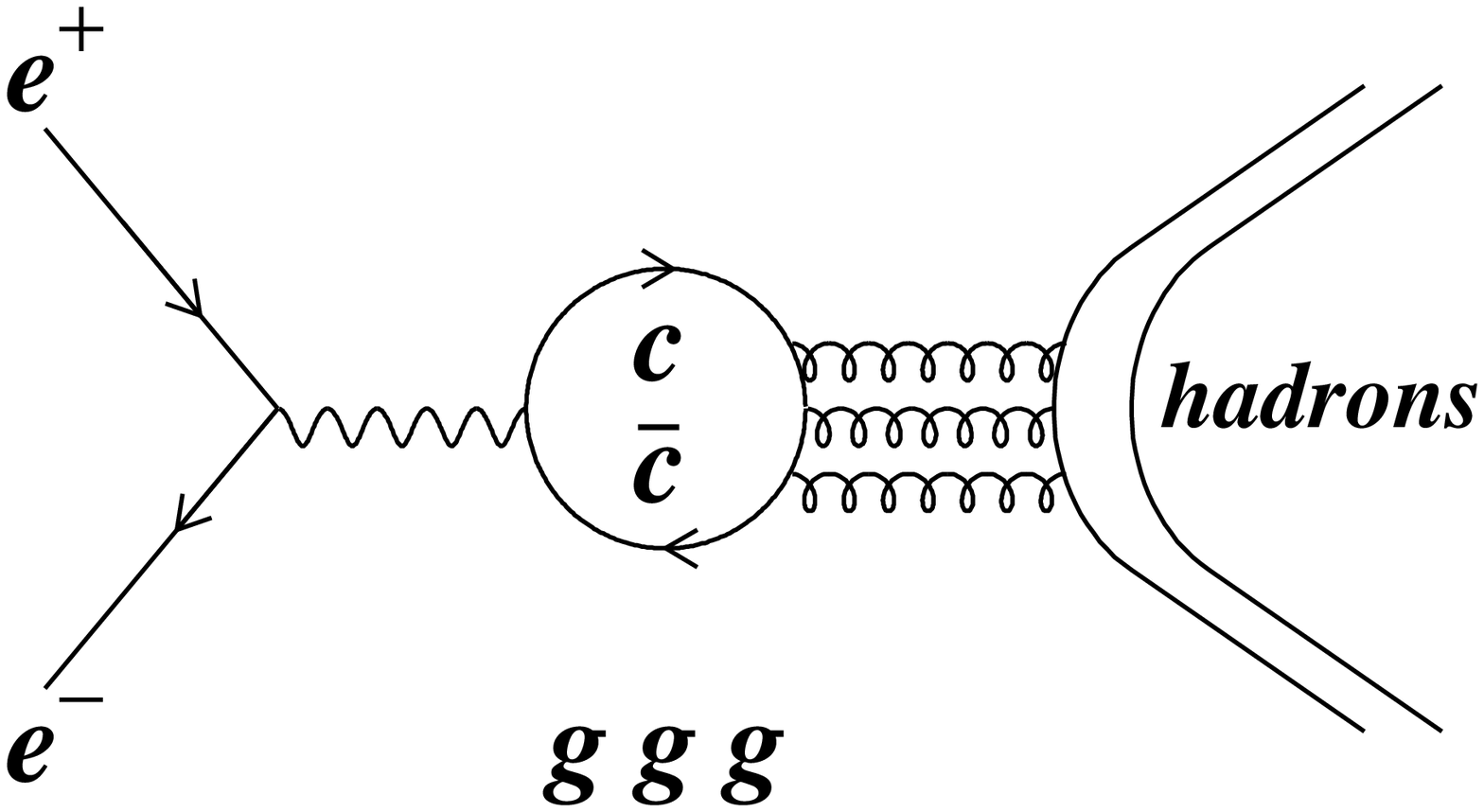}
\hskip 0.5cm
\includegraphics[width=3.25cm,height=2.5cm]{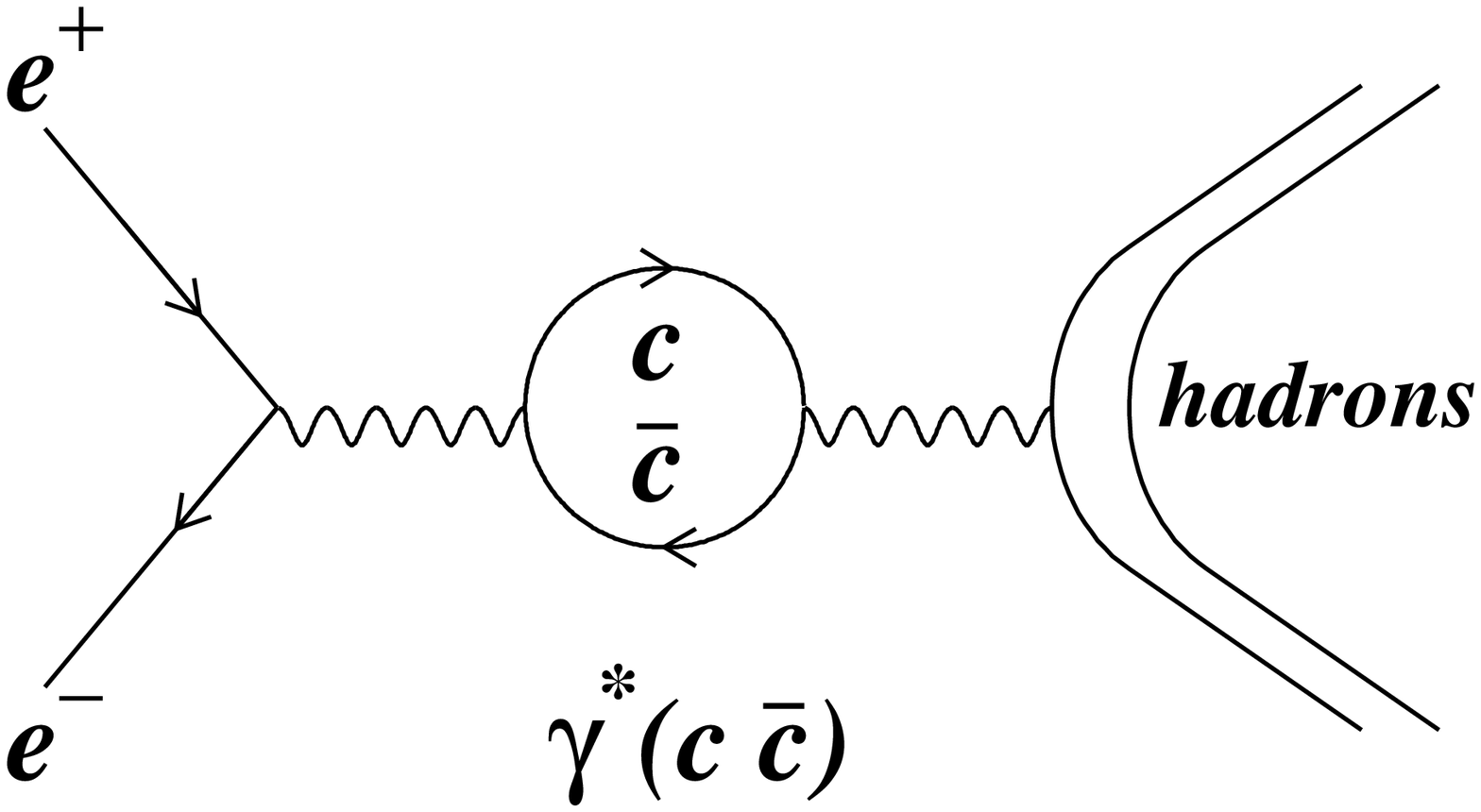}
\end{minipage}
\begin{minipage}{8cm}
\center
\includegraphics[width=3.5cm,height=2.5cm]{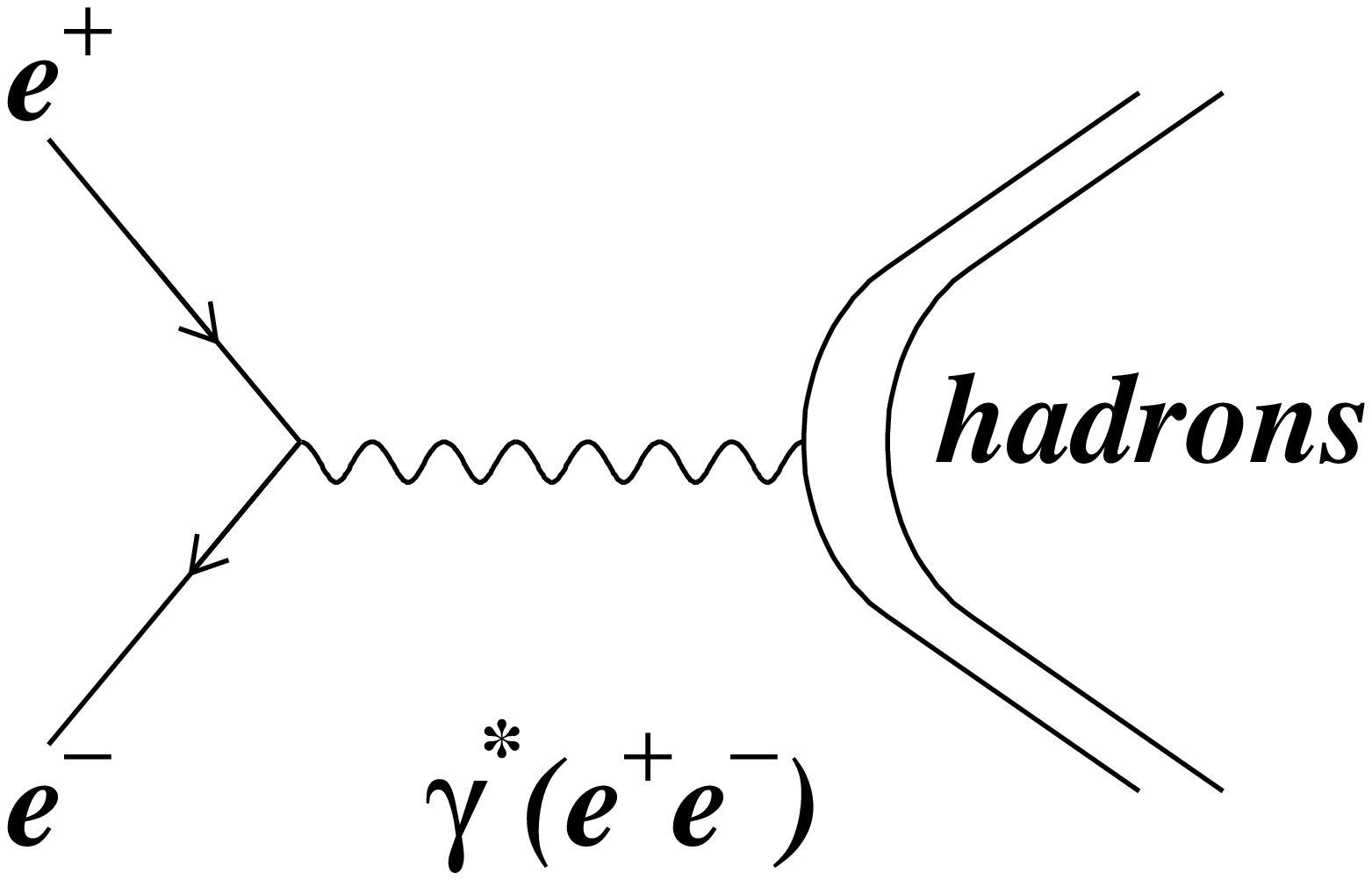}
\end{minipage}
\caption{\label{threefymn} The three classes of diagrams of
$\EE\rightarrow light\, \, hadrons$ at charmonium resonance. The
charmonium state is represented by a charm quark loop.}
\end{figure}

The OZI suppressed decays of the $\pspp$ into light hadrons are via
strong and electromagnetic interactions. In general, the cross
section of $\EE$ to a certain final state at the resonance is
expressed in the Born order by
 \beq
 \sigma_B(s)= \frac{4\pi s\alpha^2}{3}|\ag(s)+\aga(s)|^2{\cal P}(s).
 \label{born}
 \eeq
In the above equation, $\sqrt{s}$ is the center-of-mass (C.M.)
energy of $\EE$, $\alpha$ is the QED fine structure constant,
$\ag(s)$ and $\aga(s)$ denote the amplitudes in which the
resonance decays via strong and electromagnetic interactions
respectively, ${\cal P}(s)$ is the phase space factor for the
final states. However, in $\EE$ colliding experiments, the
continuum process
 \begin{equation}
 e^+e^- \rightarrow \gamma^* \rightarrow hadrons \nonumber
 \end{equation}
may produce the same final hadronic states as the resonance decays
do. We denote its amplitude by $\ac$, then the cross section
becomes~\cite{rudaz,wymz}
 \beq
 \sigma_{B}(s) =\frac{4\pi s \alpha^2}{3}
 |\ag(s)+\aga(s)+\ac(s)|^2~{\cal P}(s). \label{bornp}
 \eeq
So what truly contribute to the experimentally measured cross
section are three classes of diagrams, $i.e.$ the strong
interaction presumably through three-gluon annihilation, the
electromagnetic interaction through the annihilation of
$c\overline{c}$ pair into a virtual photon, and the one-photon
continuum process, as illustrated in Fig.~\ref{threefymn}, where
the charm loops stand for the charmonium state. To analyze the
experimental results, we must take these three amplitudes into
account.

For an exclusive mode, $\ac$ can be expressed by
 \begin{equation}
 \ac(s) = \frac{{\cal F}(s)}{s},
 \label{ac}
 \end{equation}
where ${\cal F}(s)$ is the electromagnetic
form factor which may have a dimension
depending on the final states, here we define ${\cal F}(s){\cal
P}(s)$ to be dimensionless. We adopt the convention that $\ac$ is
real. Since $\aga$ is due to the resonance, it is expressed in the
Breit-Wigner form
 \begin{equation}
 \aga(s)=\frac{3\Gamma_{ee}{\cal F}(s)/(\alpha \sqrt{s})}
 {s-M^2+i M \Gamma_t},
 \label{aga}
 \end{equation}
where $M$ and $\Gamma_t$ are the mass and the total width of the
resonance, $\Gamma_{ee}$ is its partial width to $\EE$. As $\ag$
is also due to the resonance decays, it can be parametrized
relative to $\aga$ by a complex factor
 \beq {\cal C}\equiv |\ag/\aga|e^{i \phi},
 \label{c}
 \eeq
as
 \begin{equation}
 \ag(s) = {\cal C} \cdot \frac{3\Gamma_{ee}{\cal F}(s)/(\alpha \sqrt{s})}
 {s-M^2+i M \Gamma_t}~.
 \label{ag}
 \end{equation}

Neglecting double OZI suppressed processes, 
the amplitudes of $\pspp$ decays into pairs of vector-pseudoscalar
mesons ($VP$) are parametrized in terms of three parameters: the
strong amplitude $g$, the electromagnetic amplitude $e$ and the
$SU(3)$ symmetry breaking factor $(1-s_g)$~\cite{phyreport,haber}.
These are listed in Table~\ref{tcleo} for some of the measured
decay modes together with the measured values by 
CLEOc~\cite{cleolp05} and BESII~\cite{besvp,besrp}. Similar
parametrization also applies to other OZI suppressed two-body
decays which conserve the generalized C-parity  by appropriate
change of labelling~\cite{phyreport}. As for the decays to pairs
of pseudoscalar-pseudoscalar mesons ($PP$) which violate the generalized
C-parity, the amplitudes are parametrized in terms of two
parameters: the electromagnetic amplitude $E$ and strong $SU(3)$
breaking amplitude ${\cal M}$~\cite{phyreport,haber}. These are
listed in Table~\ref{pp}, together with the BESII and CLEOc
measurements~\cite{bes2pspkskl,cleoformf}. Similar parametrization
can be extended to pairs of vector-vector mesons ($VV$). The amplitudes
$\ag$ and $\aga$ can be expressed in terms of this parametrization
scheme. For example, for $\rhopi$ mode, $\ag=g$, $\aga=e$, $|{\cal
C}|=|g/e|$; for $\KKSN$, $\ag=g(1-s_g)$, $\aga=-2e$, $|{\cal
C}|=|g(1-s_g)/(2e)|$; while for $\KK$,
$\ag=\frac{\sqrt{3}}{2}{\cal M}$, $\aga=E$. $|{\cal
C}|=\frac{\sqrt{3}}{2}|{\cal M}/E|$. For $VP$ final states, and
other final states conserving the generalized C-parity, define
 $$
 \theta_g = \mbox{arg} \left(\frac{g}{e}\right),
 $$
while for $PP$ and $VV$ final states, define
 $$
 \theta_g = \mbox{arg} \left(\frac{{\cal M}}{E}\right),
 $$
we have $\phi=\theta_g$ if in Table~\ref{tcleo} the sign between
$g$ and $e$ is positive; and $\phi=\theta_g+180^\circ$ if the sign
between $g$ and $e$ is negative; while $\phi=\theta_g$ for $\KK$
in Table~\ref{pp}. To include $\ac$ in the formula, simply replace
$\aga$ with $(\aga+\ac)$.

\begin{table*}
\caption{\label{tcleo} The experimental results from
CLEOc~\cite{cleolp05} and BESII~\cite{besvp,besrp} on $\pspp$
decays into $VP$ modes and $b_1\pi$. Also listed is the
parametrization of their amplitudes. Neglecting double OZI
suppressed processes, the amplitudes of $\pspp$ decays into $VP$
final states are parametrized by three terms: the strong amplitude
$g$, the electromagnetic amplitude $e$ and the SU(3) breaking
factor $(1-s_g)$~\cite{phyreport,haber}. 
Similar parametrization can be extended to other
OZI suppressed two-body decays which conserve the generalized C-parity
by appropriate change of labelling. The physical states $\eta$ and
$\eta^\prime$ are expressed by their quark components as $|\eta
\rangle = X_\eta \frac{1}{2} |u\overline{u} + d\overline{d}
\rangle + Y_\eta |s \overline{s} \rangle$ and $|\eta^\prime
\rangle = X_{\eta^\prime} \frac{1}{2} |u\overline{u} +
d\overline{d} \rangle + Y_{\eta^\prime} |s \overline{s} \rangle$.}
\begin{ruledtabular}
\begin{tabular}{cccccc}
Channel & Amplitude & \multicolumn{2}{c}{CLEOc }
                    & \multicolumn{2}{c}{BESII}             \\
        &           & $\sigma$(3.671 GeV) [pb]
                              & $\sigma$(3.773 GeV) [pb]
                    & $\sigma$(3.650 GeV) [pb]
                              & $\sigma$(3.773 GeV) [pb]   \\ \hline
%%%%%%%%%%%%%%%%%%%%%%%%%%%%%%%%%%%%%%%%%%%%%%%%%%%%%%%%%%%%%%%%%%%%%%%%%%%%
\multicolumn{6}{c}{VP }                                   \\ \hline
$\rho^+ \pi^-,~ \rho^- \pi^0, ~\rho^- \pi^+$
        &$g + e$& $8.0^{+1.7}_{-1.4}\pm 0.9$
                      & $4.4 \pm 0.3 \pm 0.5$
        & $< 25$              & $<6.0$                     \\
$\OP$   & $ 3e$   & $15.2^{+2.8}_{-2.4}\pm 1.5$
                      & $14.6 \pm 0.6 \pm 1.5$
    & $24.2^{+11}_{-9}\pm 4.3$& $10.7^{+5.0}_{-4.1}\pm 1.7$ \\
$\phi \pi^0$
        & $ 0$      & $< 2.2$ & $< 0.2$                 & &  \\
$\RET$  & $3e X_{\eta}$
                    & $10.0^{+2.2}_{-1.9}\pm 1.0$
                          & $10.3 \pm 0.5 \pm 1.0$
       & $8.1^{+7.4}_{-4.9}\pm 1.1$& $7.8^{+4.4}_{-3.5}\pm 0.08$ \\
$\OET$  & $(g + e) X_{\eta}$
                    & $2.3^{+1.8}_{-1.1}\pm 0.5$
                          & $0.4^{+0.2}_{-0.2} \pm 0.1$ & & \\
$\FET$  & $[g(1-2s_{g}) -2e] Y_{\eta}$
                    & $ 2.1^{+1.9}_{-1.2} \pm 0.2$
                              & $ 4.5 \pm 0.5 \pm 0.5$    & &\\
$\RETp$ & $3e X_{\eta^{\prime}}$
                    & $2.1^{+4.7}_{-1.6}\pm 0.2$
                          & $3.8^{+0.9}_{-0.8}\pm 0.4$
        & $<89 $              & $<28$                          \\
$\OETp$ & $(g + e) X_{\eta^{\prime}}$
                   & $<17.1$  & $0.6^{+0.8}_{-0.3}\pm 0.6$ & & \\
$\FETp$ & $[g(1-2s_{g}) - 2e] Y_{\eta^{\prime}}$
                   & $<12.6$
                              & $2.5^{+1.5}_{-1.1} \pm 0.4$ & & \\
$K^{*0}\overline{K^0},~\overline{K}^{*0} K^0$
        & $g(1-s_{g}) - 2e $
                   & $23.5^{+4.6}_{-3.9}\pm 3.1$
                      &$23.5 \pm 1.1 \pm 3.1$   & & \\
$K^{*+} K^- , ~K^{*-} K^+$
        & $g(1-s_{g}) + e $
                   & $ 1.0^{+1.1}_{-0.7} \pm 0.5$
                              &$ <0.6 $               & &  \\ \hline
\multicolumn{6}{c}{AP }                                 \\ \hline
$b_1\pi$&$g + e$& $7.9^{+3.1}_{-2.5}\pm 1.8$
                      & $6.3 \pm 0.7 \pm 1.5$  & & \\
\end{tabular}
\end{ruledtabular}
\end{table*}

\begin{table}
\caption{\label{pp} The measured 
%CLEOc\cite{cleoformf} for $\PP,\KK$ (for $K_S K_L$, the upper
%limit in parentheses is from BES~\cite{besvp} measured at $E_{cm}
%=3.65$~GeV). 
$\PP,\KK$ cross sections at $E_{cm}=3.671$~GeV 
by  CLEOc\cite{cleoformf}
and the upper limit of $K_S K_L$ cross sections 
at $E_{cm}=3.65$~GeV by BES~\cite{bes2pspkskl}.
Also listed is the parametrization of $\pspp$ decays
into $PP$ final states. These amplitudes are parametrized in terms
of electromagnetic amplitude $E$ and strong $SU(3)$ breaking
amplitude ${\cal M}$~\cite{phyreport,haber}. 
Similar parametrization can be applied to
$VV$ final states.}
\begin{ruledtabular}
\begin{tabular}{ccc}
final state & $\sigma$ (3.671~GeV) & amplitude \\ \hline
 $\PP$  &  $ 9.0\pm 1.8\pm 1.3$~pb & $E$  \\
 $\KK$  &  $ 5.7\pm 0.7\pm 0.3$~pb & $E+\frac{\sqrt{3}}{2}{\cal M}$ \\
$\kskl$ & ($ <5.9$~pb at 90\% C.L. at 3.65~GeV)
                                   & $\frac{\sqrt{3}}{2}{\cal M}$
\\ \hline
\end{tabular}
\end{ruledtabular}
%\end{table*}
\end{table}

If both the strong and electromagnetic interactions exist, the
partial width of $\pspp \ra f$ is calculated by
 \beq \Gamma_f =\Gamma_{ee}|{\cal F}(s)|^2\cdot|1+{\cal C}|^2{\cal P}(s).
 \label{gammaf}
 \eeq
If the decay to the final state only goes via electromagnetic
interaction, then
 \beq
 \Gamma_f = \Gamma_{ee}|{\cal F}(s)|^2\cdot{\cal P}(s).
 \label{gammafe}
 \eeq
As for the decay to $\kskl$ mode which only goes via strong
interaction (here we assume that there is no excited $\phi$ states
in the vicinity), ${\cal F}(s)=0$ since this mode does not couple
to a virtual photon. Nevertheless, according to the
parametrization in Table~\ref{pp}, the decay modes $\KK$ and
$\kskl$ have the same strong decay amplitude, while $\PP$ and
$\KK$ have the same electromagnetic decay amplitude, which is
consistent with the recently measured data listed in
Table~\ref{pp} within the experimental errors~\cite{cleoformf}.
Under such assumption, $|{\cal C}|$ can be determined by the
equation
 \beq
 \Gamma_{\kskl} = \Gamma_{ee}|{\cal F}_{\pp}(s)|^2
 \cdot|{\cal C}|^2{\cal P}(s). \label{gammafs}
 \eeq
Then with the measurement of $\KK$ mode, $\phi$ is obtained.

If the data is taken at the energy of the $\pspp$ mass, i.e. $s=M^2$,
from Eq.~(\ref{aga}) we see that $\aga$ has a phase of $-90^\circ$
relative to $\ac$. So there is no interference between $\ac$ and
$\aga$. For the decay modes which go only via electromagnetic
interaction, e.g. $\omega\pi^0$, $\rho\eta$, $\rho\eta^\prime$ and
$\pi^+\pi^-$, the interference between the resonance and continuum
can be neglected. (Under such circumstance, the interference is
still non-vanishing due to two reasons: first, for the practical
reason in the experiments, the data is usually taken at where the
maximum inclusive hadron cross section is, which in general does
not coincide with the mass of the resonance~\cite{Zline}; second,
even if the data is collected at the energy of the resonance mass,
the interference is non-vanishing because of radiative correction.
This will be proved later on in this paper. For the narrow
resonances with widthes smaller than the energy spread of the
$\EE$ colliders, the smearing of the C.M. energy also results in a
non-vanishing interference term. But these are beyond the concern
for the accuracy of current experiments.) Under such circumstance,
in the data analysis we simply subtract the continuum cross
section from the cross section measured on top of the resonance to
get the resonance cross section. The ratio of the resonance cross
section of a particular final state to the total resonance cross
section gives the branching fraction of this mode. For the $\pspp$,
$\aga$ is very small compared to $\ac$. This is seen that if
$s=M^2$,
 $$
 \left| \aga(M^2_{\pspp})/\ac(M^2_{\pspp}) \right|
 = \frac{3}{\alpha} \BR(\pspp \ra \EE).
 $$
With the measured value of $\BR(\pspp \ra \EE) = (1.12 \pm 0.17)
\times 10^{-5}$~\cite{pdg},
 $$
 \left| \aga(M^2_{\pspp})/\ac(M^2_{\pspp})
 \right| \approx 4.6 \times 10^{-3}.
 $$
So $\aga$ can be neglected. For those modes which only go via
electromagnetic interaction, the measured cross sections at the
$\pspp$ mass almost entirely come from the non-resonance continuum
amplitude $\ac$. This is demonstrated by the experimental results
on $\omega\pi^0$, $\rho\eta$, and $\rho\eta^\prime$ modes in
Table~\ref{tcleo} where their cross sections measured at the
$\pspp$ peak are consistent with the ones measured off the
resonance within experimental errors with the later scaled for $s$
dependence.

But for other final states which have contributions from $\ag$
besides $\aga$, there could be interference between $\ag$ and
$\ac$ as well as between $\ag$ and $\aga$. Since for the $\pspp$,
$\aga$ is very small compared to $\ac$, so only the interference
between $\ag$ and $\ac$ could be important. Based on the analysis
of the experimental data, we have suggested that the phase
$\theta_g$ is universally $-90^\circ$ in quarkonium
decays~\cite{wym4,wym2}. Since at the energy of resonance mass,
the phase of $\aga$ is $-90^\circ$ relative to $\ac$, so the
relative phase between $\ag$ and $\ac$ is either $180^\circ$ or
$0^\circ$, depending on whether the relative sign between $g$ and
$e$ in Table~\ref{tcleo}, or between ${\cal M}$ and $E$ in
Table~\ref{pp} is plus or minus. The interference between $\ag$
and $\ac$ is destructive for the final states $\rho\pi$,
$\omega\eta$, $\omega\eta^\prime$, $\KKSC + c.c.$, $b_1\pi$, and
$\KK$, but constructive for $\phi\eta$, $\phi\eta^\prime$, and
$\KKSN + c.c.$

Destructive interference between $\ag$ and $\ac$ means that the
observed cross section at the energy of the resonance mass can be
smaller than the continuum cross section. The experimental results
on $\rho\pi$ and $\omega\eta$ modes in Table~\ref{tcleo}
demonstrate this interference pattern.

\section{Radiatively corrected cross section}
\label{sec_formf}

The actual description of $e^+e^-$ to a final hadronic state
through the annihilation of a virtual photon must incorporate
radiative correction. Such correction mainly comes from the
initial state radiation, and for the hadronic final state, the
final state radiation usually can be neglected~\cite{ptsai}.

The integrated cross section by $\EE$ collision incorporating
radiative correction is expressed by~\cite{rad}
 \begin{equation}
 \sigma_{r.c.} (s)=\int \limits_{0}^{1-s_m/s} dx F(x,s)
 \sigma_{B}(s(1-x))
 \label{radsec}
 \end{equation}
where
$\sigma_{B}(s)$ is the Born order cross section by
Eq.~(\ref{bornp}),
and in the upper limit of the integration $\sqrt{s_m}$ is the
cut-off invariant mass of the final state
hadron system after losing energy to photon emission. $F(x,s)$ is
calculated to an accuracy of $0.1\%$ in Ref.~\cite{rad}:
\begin{widetext}
\beqn
 F(x,s)&=&
 \beta x^{\beta-1} \left[1+\frac{3}{4}\beta+\frac{\alpha}{\pi}
 \left(\frac{\pi^{2}}{3}-\frac{1}{2}\right)+\beta^{2}
 \left(\frac{9}{32}-\frac{\pi^{2}}{12}\right) \right]
 -\beta\left(1-\frac{x}{2}\right) \nonumber \\
 &+ &\frac{1}{8}\beta^{2} \left[4(2-x)\ln\frac{1}{x}-
 \frac{1+3(1-x)^{2}}{x}\ln(1-x)-6+x\right],
 \label{Fexp}
 \eeqn
\end{widetext}
with
 \begin{equation}
 \beta\;=\;\frac{2\alpha}{\pi} \left(\ln
 \frac{s}{m^{2}_{e}}-1\right).
 \label{beta}
 \end {equation}

In the above equations, $m_e$ is the mass of the electron. Here
the expression $F(x,s)$ includes the bremsstrahlung of an $\EE$
pair from the initial $\EE$ state. For $x \sim 0$, $x \sqrt{s}/2$ is
approximately equal to the energy carried away by the radiated
photons. But for $x \sim 1$, this meaning is not valid in the $\alpha^2$
order. The effects of vacuum polarization are not included explicitly in
Eq.~\eref{radsec}. Here we follow the convention that for hadronic
final state, the vacuum polarization by leptons and hadrons,
including vector-meson resonances, is taken into account in the
form factor ${\cal F}(s)$~\cite{Greco}.

In Eq.~\eref{radsec}, $F(x,s)$ is positive definite. If $s=M^2$,
for pure electromagnetic processes, the interference term between
$\aga$ and $\ac$ in $\sigma_{B}(s(1-x))$, i.e.
\begin{widetext}
\beqn
 2 \Re \left[\aga(s(1-x)) \ac(s(1-x)) \right]
&= & 2 \Re \left[
 \frac{{\cal F}(s(1-x))}{s(1-x)} \cdot \frac{3\Gamma_{ee}{\cal F}
 (s(1-x))/(\alpha \sqrt{s(1-x)})}
 {s(1-x)-M^2+iM\Gamma_t} \right] \nonumber \\
 & = & 2 \left[ s(1-x) - M^2 \right] 
 \cdot \frac{3 \left|{\cal F}(s(1-x)) \right|^2 \Gamma_{ee}}
 {\alpha [s(1-x)]^{3/2} \left[ (s(1-x) - M^2)^2 + M^2\Gamma_{t}^2\right]}~,
 \eeqn
\end{widetext}
vanishes only at $x=0$, but it is negative elsewhere in the
integration interval from 0 to $1-s_m/s$. So even in the pure
electromagnetic processes, with radiative correction, the
interference term gets a negative value.

In Eqs.~\eref{ac}, \eref{aga} and \eref{ag}, the form factor
${\cal F}(s)$ is adopted to describe the hadronic interaction.
Both the Monte Carlo simulation and the calculation of the
radiatively corrected cross section require the knowledge of the
form factor in the energy range from $\sqrt{s}$ down to
$\sqrt{s_m}$. In principle, $\sqrt{s_m}$ can be as low as the
production threshold. But this requires the input values of the
form factor from $\sqrt{s}$ to the production threshold. In
Eq.~\eref{radsec}, $F(x,s)$ has been calculated to an accuracy of
$0.1\%$ by QED, but for virtually all the hadronic final states,
we have neither the theoretical models nor sufficient experimental
data to describe the form factors to such high precision. To
reduce the systematic uncertainty from the Monte Carlo
simulation, one strategy is to take the value of $s_m$ as %large as
close to $s$ as possible for the actual event selection criteria, 
i.e. we generate the event sample with $s_m$ just lower than that which 
are to be selected. In this way, we need only to describe the form factor
precisely from $\sqrt{s}$ down to a much higher $\sqrt{s_m}$,
instead of to the production threshold.

The variation of the form factor 
in a small energy interval usually can well be
approximated by
 \beq
 {\cal F}(s^\prime)={\cal F}(s)\left(\frac{s}{s^\prime}\right)^k,
 \label{formf}
 \eeq
with $k$ either derived from theoretical models, or obtained from
fitting the experimental data at nearby energy. Of course one may
use a more complicated function other than Eq.~(\ref{formf}) for
the approximation.

\section{Monte Carlo simulation}

\subsection{Soft and hard photon events}

In realization of Monte Carlo simulation, an auxiliary
parameter $x_0$ is introduced to separate the integration interval
of Eq.~\eref{radsec} into two parts, $(0, x_0)$ and $(x_0,
1-s_m/s)$, with $x_0$ a small but nonzero value:
 \beq
 \sigma_{soft}(s) = \int \limits_{0}^{x_0} dx
%F(x,s) \frac{\sigma_{B}(s(1-x))}{|1-\Pi (s(1-x))|^2}
F(x,s) \sigma_{B}(s(1-x))
 \label{xsoft}
 \eeq
and
 \beq
 \sigma_{hard} (s) = \int \limits_{x_0}^{1-s_m/s} dx
 %F(x,s) \frac{\sigma_{B}(s(1-x))}{|1-\Pi (s(1-x))|^2}~.
 F(x,s) \sigma_{B}(s(1-x))
 \label{xhard}
 \eeq
In the above two equations, all functions and variables have the
same meaning as in Eq.~\eref{radsec}. We have
$$\sigma_{r.c.}(s)=\sigma_{soft}(s)+\sigma_{hard}(s)~.$$
The two terms in the above expression are usually called soft
photon and hard photon cross sections respectively. The Monte
Carlo program generates the soft photon events and hard photon
events according to their proportions in the total cross section.
In a soft photon event, the four-momentum of the radiated photons
are neglected, and the photons are not generated. Experimentally,
this means that in the soft photon events, the photons can not be
detected. Since in Eq.~\eref{xsoft}, $x_0$ is small, so
$x_0\sqrt{s}/2$ can be identified as the maximum energy carried
away by the undetected photons. This requires that $x_0\sqrt{s}/2$
must be smaller than the minimum energy of the photon which can be
detected in the experiment. In a soft photon event, the
energy-momentum is conserved between the incoming $\EE$ pair and
the final state hadron system, so $x_0$ must also be smaller than
the energy-momentum resolution of the detector. Usually $x_0$ is
assigned a nonzero value smaller than $0.01$. The outcome of the
Monte Carlo simulation does not depend on $x_0$. In a hard photon
event, the four-momentum of the radiated photons are generated,
and the energy-momentum is conserved with the inclusion of these
photons.

The differential cross sections with the emission of photons for
exclusive hadronic processes are calculated in
Refs.~\cite{hard1,hard2}, the calculation of Ref.~\cite{ISR} can
also be used to generate the hard photon events while the
generation of soft photon events is identical to the generation of
events at Born order. (The inclusive process is treated somewhat
differently in Ref.~\cite{jadach}.) The authors of these
references also provide Monte Carlo programs based on their
calculations. The program based on the calculation in
Ref.~\cite{hard1}, BABAYAGA, generates $\pi^+\pi^-$ events; the
program based on Ref.~\cite{hard2}, MCGPJ with the current
version, generates $\pi^+\pi^-$, $K^+K^-$, and $K_SK_L$ events;
while the program based on Ref.~\cite{ISR} generates hard photon
events for $\pi^+\pi^-$, $\pi^+\pi^-\pi^0$,
$\pi^+\pi^-\pi^+\pi^-$, and $\pi^+\pi^-\pi^0\pi^0$ processes.
These programs achieve the precision of $(0.1 \sim 0.2)\%$.

In order to generate events with the presence of both the $\pspp$
resonance and continuum, some replacements are to be made in the
fore-mentioned programs. The Born order cross section by
Eq.~\eref{bornp} must be substituted to generate the correct
distribution of the invariant mass of the final state hadron
systems. Although the original programs only generate a few
hadronic final states, the programs based on
Refs.~\cite{hard2,ISR} are written in the form that more final
states can be added in a straightforward way. To do this, one need
to put the corresponding hadronic tensor $H_{\mu\nu}$ into the
program, where \beq H_{\mu\nu} = {\cal H}_{\mu} \times {\cal
H}^{*}_{\nu}, \eeq with ${\cal H}_{\mu}$ the current of virtual
photon transition to the final hadron state $f$. For reference we 
include its forms for some final hadronic states in the appendix.

\subsection{Distribution of invariant mass of hadrons}

\label{Mdistribution}

The distribution of the invariant mass of the final hadron system
$M_{inv}$ is very different between resonance and continuum. More
profound distinctive feature is in the circumstance that there is
interference, particularly destructive one, between the resonance and
continuum. In this section, we discuss this distribution.

\subsubsection{Resonance}

If the final state does not couple to a virtual photon, 
e.g. $\kskl$,
then in Eq.~\eref{bornp}, there is only the $\ag$ term, the Born
order cross section is expressed by the Breit-Wigner formula:
 \beq
 \sigma_{res}(s)=\frac{12\pi\Gamma_{ee}\Gamma_{f}}
 {(s-M^2)^2+M^2\Gamma_t^2}, \label{bw}
 \eeq
where $\Gamma_{f}$ is the partial width to the final state $f$.

For narrow resonances, by narrow we mean $\Gamma_t/M \ll x_0$,
Eq.~\eref{bw} behaves almost like a $\delta$ function.
%\beq
%\sigma_{res}(s) \approx
%\frac{12\pi^2}{M} \Gamma_{ee} \delta(s-M^2).
%\eeq
If the data is taken at the energy of the resonance mass, i.e.
$s=M^2$, in Eq.~\eref{bw}, $\sigma_{res}(s(1-x))$ is very large at
$x=0$ and very small elsewhere. So in Eq.~\eref{radsec}, the
contribution of $\sigma_B(s(1-x)$ to the radiatively corrected cross
section comes solely from a small interval around $x=0$.
Substitute $\sigma_{res}(s(1-x))$  for $\sigma_B(s(1-x))$ in
Eqs.~\eref{xsoft} and \eref{xhard}, $\sigma_{soft}$ is nonzero
while $\sigma_{hard}$ virtually vanishes due to their integration
intervals respectively. This means that in the Monte Carlo
simulation of a narrow resonance, only the soft photon events are
generated. The sole effect of the radiative correction is the
reduction of its height. This will be discussed later in this
paper.

The resonance $\pspp$ is not very narrow in this sense. If we take
$x_0=0.01$ and $s_m=0.8M^2$, soft photon events are 98.6\% of the
total.

\subsubsection{Non-resonance continuum}

For continuum process, there is a considerable  hard photon cross
section. Take an example of $VP$ final states, with $x_0=0.01$ and
$s_m=0.8s$, and assuming that the form factor varies as a function
of energy according to $1/s$ ($k=1$ in Eq.~\eref{formf}), in the
simulation the hard photon events are 22\% of the total.

\subsubsection{Interference between resonance and continuum}
\label{xint}

If both the resonance and continuum amplitudes exist, the Born
order cross section is obtained from Eq.~\eref{bornp} with $\ac$,
$\aga$ and $\ag$ by Eqs.~\eref{ac}, \eref{aga}, and \eref{ag}.
\begin{widetext}
\beqn
 \sigma_{B} (s(1-x))
 & = & \frac{4\pi s(1-x) \alpha^2}{3}  \left|\ac(s(1-x)) +
       \ag(s(1-x)) + \aga(s(1-x)) \right|^2 {\cal P}(s(1-x))
    \nonumber \\
 & = & \frac{4}{3}\pi s(1-x) \alpha^2|{\cal F}(s(1-x))|^2
       {\cal P}(s(1-x))\times \left|\frac{1}{s(1-x)} +
({\cal C}+1) \frac{3\Gamma_{ee}/\left(\alpha \sqrt{s(1-x)}\right)}
 {s(1-x)-M^2 + i M\Gamma_{t}} \right|^2.
\label{int}
\eeqn
\end{widetext}

We pay special attention to the destructive interference and take 
as an extreme exmaple the circumstance that $\ag$ and $\ac$
almost cancel to each other in the Born order. This happens if
 \beq
 |{\cal C}| \approx \frac{\alpha \Gamma_t}{3\Gamma_{ee}},~~~
 \phi=-90^\circ. \label{cancel}
 \eeq
Under such circumstance, the partial width of this final state is
expressed by
 \beq
 \Gamma_f \approx \frac{M^2\Gamma_t^2}{12\pi\Gamma_{ee}} \sigma_{con}(M^2),
 \eeq
where $\sigma_{con}(M^2)$, scaled for $s$ dependence to $s=M^2$, is the 
Born order cross section for the non-resonance continuum.

To calculate the proportion of soft and hard photon events,
Eq.~\eref{int} is to be substituted into Eqs.~\eref{xsoft} and
\eref{xhard} respectively. Notice that these two equations only
differ by their integration intervals. For the discussions on
the $\pspp$ in this work, we take $x_0=0.01$ which is greater than
$\Gamma_t/M$. If the resonance amplitude satisfies
Eq.~\eref{cancel}, then for Eq.~\eref{xsoft}, in the integration
interval $(0, x_0)$, there is almost complete cancellation between
the two terms of $\sigma_{B}(s(1-x))$; while for Eq.~\eref{xhard},
in the integration interval $(x_0, 1-s_m/s)$, the magnitude of
the resonance
which is the second term of $\sigma_{B}(s(1-x))$, virtually
vanishes. This means that the resonance and its interference with
continuum affect mainly the soft photon cross section; while the
hard photon cross section is predominately due to $\ac$. In this
way, the interference  changes the proportions of the soft and
hard photon events as well as the total radiatively corrected cross
section. The destructive interference between resonance and
non-resonance continuum reduces the soft photon cross section,
which means a smaller proportion of soft photon events.

To illustrate the above discussion quantitatively, we take the
example that the resonance amplitude satisfies Eq.~\eref{cancel}.
Under such circumstance, the proportion of the soft photon events
is 15\% for $x_0=0.01$ and $s_m=0.8s$. (This is not the lowest
possible proportion of the soft photon events. With $\phi =
-90^\circ$, the complete cancellation happens only between $\ag$
and $\ac$. There is still a small but non-vanishing $\aga$ left.
If $\ag$ cancels out not only $\ac$, but also $\aga$, the
proportion of soft photon events can be even smaller.) This means
that for a detector with energy-momentum resolution $x_0/2=0.5\%$,
among the data taken at non-resonance continuum, 78\% of the
events have invariant mass equal to the C.M. energy of the
incoming $\EE$; while among the data taken at the $\pspp$ mass, we
may find that the majority of the events (85\%) have invariant
mass smaller than the C.M. energy of the incoming $\EE$. If this
happens, it indicates destructive interference between the resonance
and non-resonance continuum amplitudes.

\begin{figure}[htbp]
\includegraphics[width=7.cm,height=6.0cm]{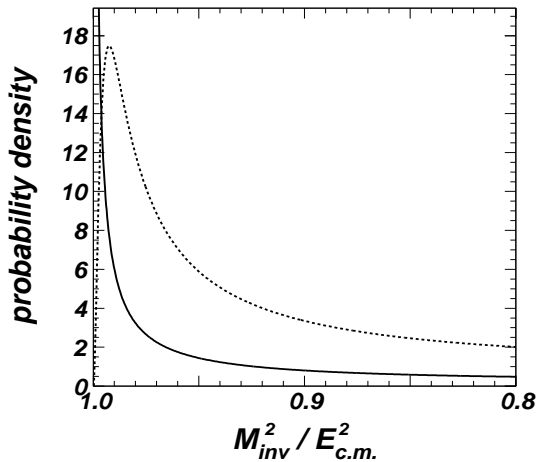}
\caption{\label{curve1}The distribution of the probability density
as a function of $M_{inv}^2/s$ with $M_{inv}$ the invariant mass
for a $VP$ final state. The solid line is for continuum. The
dashed line is an example of destructive interference between
$\ag$ and $\ac$ with the two amplitudes satisfying
Eq.~\eref{cancel}. The probability is normalized to
$\sigma_{r.c.}(M^2_{\pspp})$ with $s_m=0.8s$.}
\end{figure}

Fig.~\ref{curve1} shows the probability density as a function of
the squared invariant mass of the final hadron system $M_{inv}^2$.
For continuum, as well as for no interference or constructive
interference between the resonance and continuum cases, the
maximum probability density occurs (actually diverges) at
$M_{inv}^2=s$, which corresponds to $E_\gamma \ra 0$ with
$E_\gamma$ the energy of the emitted photons; but if the
destructive interference between $\ag$ and $\ac$ exists, 
and it leads
to almost complete cancellation of the two amplitudes, the
probability density may have a minimum point near $M_{inv}^2=s$.

In the circumstance that at $s=M^2$, the three terms in
Eq.~\eref{bornp} completely cancel, i.e. $\ag+\aga+\ac = 0$, so
the Born order cross section vanishes, the radiatively corrected
cross section $\sigma_{r.c.}(s)$ still gets a nonzero value, as
long as $s_m < s$. This is because that in Eq.~\eref{radsec},
$F(x,s)$
is a positive definite function, while $\sigma_B(s(1-x))$ vanishes
only at $x=0$, but remains positive definite elsewhere in
the integration interval. This leads to the phenomena in $\EE$
experiments: a detector with finite energy-momentum resolution
always observes a non-vanishing cross section, even if the Born
order cross section vanishes due to destructive interference
between the resonance and continuum.

\subsection{The calculated efficiency}

According to Eq.~\eref{radsec}, $\sigma_{r.c.}$ is a function of
$s_m$ as well as $s$. For the reason which we discussed in 
Sec.~\ref{sec_formf}, in the Monte Carlo simulation, usually we generate
events to a cut-off invariant mass $\sqrt{s_m}$, so the calculated 
efficiency $\epsilon$ is also a function of $s_m$.

If the observed number of event is denoted as $N$, corresponding
integrated luminosity is denoted as ${\cal L}$, then we have
$$ N = {\cal L} \cdot \sigma_{r.c.}(s,s_m) \cdot \epsilon({s_m}),$$
here and in the following discussions, we explicitly indicate the
dependence of $\sigma_{r.c.}$ and $\epsilon$ on $s_m$. The above
equation can also be expressed as
 \beq
 \frac{N}{\cal L} = \sigma_{r.c.}(s,s_m) \cdot \epsilon({s_m}).
 \label{sigrceff}
 \eeq
The left side of the above expression is an experimentally
measured quantity. The product $\sigma_{r.c.}(s,s_m) \cdot
\epsilon({s_m})$ does not depend on $s_m$.

Very often, the experimental results are presented in terms of
Born order cross section, particularly the off resonance continuum
cross section. For the data taken off the resonance, we measure
the electromagnetic form factor of the final state, which is
simply related to the Born order cross section. A so-called
radiative correction factor is introduced as
 \beq
 f_{ISR}(s,s_m) = \frac{\sigma_{r.c.}(s,s_m)}{\sigma_{B}(s)}.
 \label{fatisr}
 \eeq
Here if the form factor takes the form of Eq.~(\ref{formf}), then
$f_{ISR}(s,s_m)$ can be calculated. It does not depend on ${\cal
F}(s)$, although it still depends on $k$ in Eq.~(\ref{formf}).
With Eqs.~\eref{sigrceff} and \eref{fatisr}, we have
$$ \sigma_{Born}(s) = \frac{\sigma_{r.c.}(s,s_m)}{f_{ISR}(s,s_m)}
=\frac{N}{{\cal L} \cdot \epsilon(s_m) \cdot f_{ISR}(s,s_m)}~.$$
In the denominator, the product of $\epsilon({s_m}) \cdot
f_{ISR}(s,s_m)$ cancels out the dependence on $s_m$.

The interference between the resonance and continuum amplitudes
may change the factor $f_{ISR}$ in a profound way. For
non-resonance continuum, with $\sqrt{s}$ well above the production
threshold, if the form factor goes down rapidly as $s$ increases
(e.g. $k \ge 1$ in Eq.~\eref{formf}), as $s_m$ approaches the
threshold, $f_{ISR}$ is usually greater than 1; but if $s_m$ is
taken close to $s$, then $f_{ISR}$ can always be smaller than 1,
since as $s_m \ra s$, $\sigma_{r.c.}(s,s_m) \ra 0$. For a
resonance, $f_{ISR}$ does not depend on $k$. It is roughly
approximated by~\cite{Zline}
 \beqn
 f_{ISR} \approx \left(\frac{\Gamma_t}{M}\right)^\beta
 \left[1+\frac{3}{4}\beta+\frac{\alpha}{\pi}
 \left(\frac{\pi^{2}}{3}-\frac{1}{2}\right)+\beta^{2}
 \left(\frac{9}{32}-\frac{\pi^{2}}{12}\right) \right].
 \eeqn
Here $\beta$ is given by Eq.~\eref{beta}. If $\Gamma_t \ll M$, the
value of this expression is less than 1. It means that the initial
state radiation reduces the height of the resonance. If there is
significant interference between the resonance and continuum,
$f_{ISR}$ may take any value between the order of 1 to infinity.
It becomes infinity in the circumstance that $\sigma_B(s)$
vanishes due to destructive interference between the resonance and
continuum. As discussed in Sec.~\ref{xint}, $\sigma_{r.c.}(s,s_m)$
still gets a nonzero value in this circumstance.

For illustrative purpose, here we take an example that in the
Monte Carlo simulation of $\EE$ collision at $\sqrt{s}=M_{\pspp}$,
the $VP$ final state events with the cut-off invariant mass of
$0.9M_{\pspp}$ are generated, and the form factor varies as a
function of the energy according to $1/s$, while in the data
selection, the invariant mass of the final $VP$ particles is
required to be greater than $98\%\sqrt{s}$. Under these
conditions, for the continuum cross section, 87.4\% of the
generated events survives and $f_{ISR}=0.946$; for the $\pspp$
resonance, $99.9\%$ of the generated events is left and
$f_{ISR}=0.716$; while for the destructive interference between
$\ag$ and $\ac$ which satisfies Eq.~\eref{cancel}, only $49.0\%$
of the generated events have invariant mass greater than
$98\%M_{\pspp}$, but $f_{ISR}=65.5$.

\section{Measurement in the presence of interference}
\label{sec_mint}

The above discussions lead to a profound feature of the
experimental measurement in the presence of interference between
the resonance and non-resonance continuum: the complete
determination of the branching fraction must come together with
the determination of the phase between the resonance and continuum
by scanned data around the resonance peak. The data must be taken
at least at four energy points, because there are three quantities
which must be determined simultaneously: ${\cal F}(M^2)$, $|{\cal
C}|$ and $\phi$. 
At the same time, the form of dependence of the observed 
cross section on $|{\cal C}|$ is quadric.
Herein
the non-resonance continuum amplitude or equivalently ${\cal
F}(M^2)$ is determined by the data taken at continuum. In the
treatment of the data taken around the resonance, if both strong
and electromagnetic interactions exist, the Monte Carlo generator
requires the input of $|{\cal C}|$ and $\phi$, so the data
analysis is an iterative process. The usual procedure is to fix
$\phi$, and varying $|{\cal C}|$, until the calculated efficiency
$\epsilon(s_m)$ by the Monte Carlo and the radiatively corrected
cross section $\sigma_{r.c.}(s,s_m)$ satisfy
Eq.~\eref{sigrceff}. Then the partial width is obtained by
Eq.~\eref{gammaf}.

If the data
are taken at two points, i.e. one at continuum off the resonance
and the other one at the energy of the resonance mass, only a 
relation between $|{\cal C}|$ and $\phi$ can be obtained. 
The solution of $|{\cal C}|$ and $\phi$ is differentiated 
into two circumstances, depending on the relative
magnitudes of the observed total cross section $\sigma_t$
and continuum cross section $\sigma_c$. Here $\sigma_c$
is the radiatively corrected cross section of the
non-resonance continuum calculated with the form factor 
at the energy $\sqrt{s}=M_{\pspp}$. 
As in the Born order cross sections, if
$\sigma_t>\sigma_c$, $\phi$ can 
take any value from $-180^{\circ}$ to $180^{\circ}$, 
and for each value of $\phi$ there is one and only one
solution for $|{\cal C}|$; 
on the other hand, if $\sigma_t < \sigma_c$, $\phi$ is
constrained within a range around $-90^{\circ}$, in 
which every possible value of $\phi$ 
corresponds to two solutions of $|{\cal C}|$.
A formal discussion is left into the appendix.
In such measurement, $|{\cal C}|$ is determined versus the
phase $\phi$ in a two dimensional curve. The recommended way is to
start from $\phi=-90^\circ$, since there is always at least one
solution of $|{\cal C}|$. If the obtained $\sigma_t$ 
is smaller than $\sigma_c$, the second solution must be searched. 
Then for the $\phi$ values greater and smaller than $-90^\circ$,
the solutions of $|{\cal C}|$ are found similarly. Thus the
curve relating $|{\cal C}|$ with $\phi$ is obtained point
by point. 

Although $\sigma_{r.c.}(s,s_m)$ and $\epsilon(s_m)$ depend on
$|{\cal C}|$ and $\phi$, for different solutions of $|{\cal C}|$
and $\phi$ which fit the experimental data, $\sigma_{r.c.}(s,s_m)$
and $\epsilon(s_m)$ only depend on $|{\cal C}|$ and $\phi$ weakly.
This can be understood by noticing that the efficiency can be
expressed in terms of the efficiency for soft photon events
$\epsilon_{soft}(x_0)$, and the one for hard photon events
$\epsilon_{hard}(s_m)$. Apparently we have
 \beq
 \sigma_{r.c.}(s,s_m) \epsilon(s_m) = \sigma_{soft}(s,x_0)
 \epsilon_{soft}(x_0) + \sigma_{hard}(s,s_m) \epsilon_{hard}(s_m).
 \label{effic}
 \eeq
As discussed in Sec.~\ref{xint}, if we take $x_0 > \Gamma_t/M$,
 the distribution of the hard photon events and the
hard photon cross section are predominately due to $\ac$, so
$\sigma_{hard}(s,s_m)$ and $\epsilon_{hard}(s_m)$ are almost
independent of $|{\cal C}|$ and $\phi$. As for the soft photon
events, the distribution follows the Born order differential cross section,
so $\epsilon_{soft}(x_0)$ does not depend on $|{\cal C}|$ and
$\phi$ either. So in Eq.~\eref{effic}, for a rough approximation,
only $\sigma_{soft}(s,x_0)$ depends on $|{\cal C}|$ and $\phi$.
The solutions of $|{\cal C}|$ and $\phi$ are to satisfy the
equation
 \beq
 \frac{N}{{\cal L}} = \sigma_{soft}(s)
 \epsilon_{soft} + \sigma_{hard}(s,s_m) \epsilon_{hard}(s_m).
 \label{fiteq}
 \eeq
Here $N/{\cal L}$ is given by the experiment, only
$\sigma_{soft}(s,x_0)$ depends on $|{\cal C}|$ and $\phi$, so
$\sigma_{soft}(s,x_0)$ remains as a constant for any possible
solution of $|{\cal C}|$ and $\phi$ which fit the data. Since
$\sigma_{r.c.}(s,s_m)=\sigma_{soft}(s,x_0)+\sigma_{hard}(s,s_m)$,
and $N/{\cal L}=\sigma_{r.c.}(s,s_m)\epsilon(s_m)$,
$\sigma_{r.c.}(s,s_m)$ and  $\epsilon(s_m)$ are almost constants
as well. This property helps to find other solutions with
different $|{\cal C}|$ and $\phi$ values once one of the solution
is found. It also makes the iterative process converge very fast.

\section{Verify the interference with the data at $\pspp$}

\begin{figure}[htbp]
\includegraphics[width=7.cm,height=6.0cm]{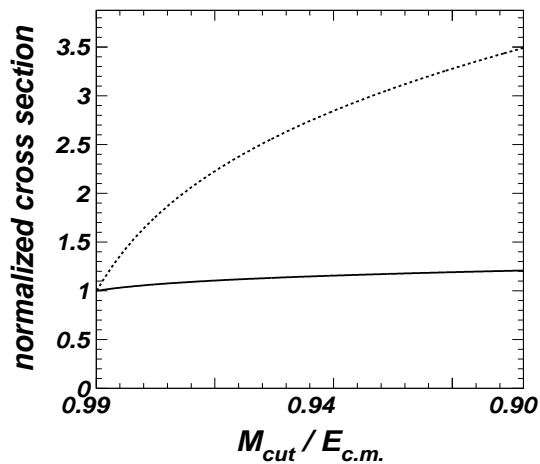}
\caption{\label{curve2}The cross section as a function of
$M_{cut}$ normalized to the cross section with
$M_{cut}=0.99\sqrt{s}$. Solid line is for continuum. Dashed line
is for destructive interference between $\ag$ and $\ac$ with the
two amplitudes satisfying Eq.~\eref{cancel}. }
\end{figure}

The discussions in Sec.~\ref{Mdistribution} leads to a scheme which
could verify the destructive interference between $\pspp$ and
non-resonance continuum with only the data at $\pspp$ peak. Modern
detectors with a CsI(Tl) calorimeter and a magnetic field of 1
Tesla or more, like CLEOc and BESIII, measure the energy-momentum
with resolution of $1\%$, which is comparable to the
ratio $\Gamma_{\pspp}/M_{\pspp}$. In this scheme the invariant
mass distribution of the hadrons is measured, and in the event
selection, requirement of the invariant mass, $M_{inv}$, greater
than a certain value, $M_{cut}$, is applied. For non-resonance
continuum, as $M_{cut}$ is loosed from $0.99\sqrt{s}$ 
to, e.g. $0.95\sqrt{s}$ or $0.90\sqrt{s}$, the
number of events increases slowly; for no interference or
constructive interference between the resonance and continuum, 
the number of events increases even slower; for pure resonance, 
the number of events does not increase at all;
on the other hand, if the destructive interference leads
to substantial cancellation between the resonance and continuum
amplitudes, as $M_{cut}$ is lowered, the number of events
increases rapidly. Table~\ref{scheme} gives the the cross section
as a function of $M_{cut}$, taking the cross section with
$M_{cut}=0.99\sqrt{s}$ as the unit. Listed cross sections are due to 
the non-resonance continuum and the destructive interference between
$\ag$ and $\ac$ with the amplitudes satisfying Eq.~\eref{cancel}.
It is assumed that the final state is $VP$ and the form factor
varies as a function of energy according to $1/s$.
Fig.~\ref{curve2} shows the cross section as a function of
$M_{cut}$ for these two circumstances, normalized to the cross
section with $M_{cut}=0.99\sqrt{s}$. In Fig.~\ref{curve2} and
Table~\ref{scheme}, we see that, as $M_{cut}$ is loosed from
$99\%\sqrt{s}$ to $90\%\sqrt{s}$, the cross section of the
continuum increases by merely $21\%$; but for the destructive
interference, it could increase by more than 3 times. In the
experimental data, if the number of events increases rapidly as
one lowers $M_{cut}$ in event selection, it indicates destructive
interference between $\pspp$ and continuum with the two amplitudes
in comparable strength.

\begin{table}
\caption{\label{scheme} The variation of the cross section as a
function of $M_{cut}$ for the non-resonance continuum and the
destructive interference between $\ag$ and $\ac$ with the two
amplitudes satisfying Eq.~\eref{cancel}, taking the cross sections
with $M_{cut}=0.99\sqrt{s}$ as the unit.}
\begin{ruledtabular}
\begin{tabular}{ccc}
${M}_{cut}/\sqrt{s}$ & destructive interference & continuum \\
\hline 0.99  & 1.00 & 1.00 \\ \hline 0.98  & 1.61 & 1.06 \\ \hline
0.97  & 2.00 & 1.09 \\ \hline 0.96  & 2.30 & 1.12 \\ \hline 0.95
& 2.53 & 1.14 \\ \hline 0.90  & 3.33 & 1.21 \\ \hline
\end{tabular}
\end{ruledtabular}
\end{table}

\begin{figure}[htbp]
\includegraphics[width=7.cm,height=6.0cm]{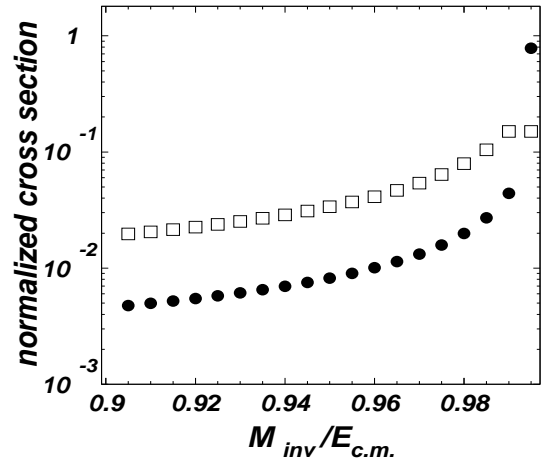}
\caption{\label{curve3}The interval of the invariant mass of the
final state hadron system from 100\% to 90\% of the $\EE$ C.M.
energy is divided into 20 bins, the probabilities of the hadron
event in each bin are plotted for continuum (dots) and for the
$\pspp$ with the destructive interference between $\ag$ and $\ac$
satisfying Eq.~\eref{cancel} (boxes).}
\end{figure}

To simulate the experimental situation, in Fig.~\ref{curve3} the
interval of the invariant mass of the final state hadron system
from 100\% to 90\% of the $\EE$ C.M. energy is divided into 20
bins, and the probabilities of the hadrons in the bins are plotted
for the events at continuum and at the energy of $\pspp$ mass with
the destructive interference between $\ag$ and $\ac$ satisfying
Eq.~\eref{cancel}. Here off resonance at continuum, the events are
highly concentrated in the last bin with $M_{inv}/E_{c.m.}=100\%$;
while for destructive interference, the events are distributed
more flatly among the bins. If there is no interference or if the
interference is constructive, the events are even more
concentrated in the last bin than off resonance at continuum in
the plot. %If there is only the resonance contribution, 
%most of the events are in the last bin.

This scheme requires the selected data sample be free from
background contamination. The most important background comes from
the radiative tail of the $\psp$. For the data taken at the $\pspp$
peak, the radiative tail due to the $\psp$ has an invariant mass which
is $97.8\%$ of the $\EE$ C.M. energy. The radiative tail of the
$\psp$ at the energy of the $\pspp$ mass has a total cross section of
2.6~nb, so this scheme is feasible for those decay modes like
$\rhopi$, $\KKSC$, and $\omega\eta$ with branching fractions in
$\psp$ decays no more than the order of
$10^{-5}$~\cite{cleopsp,bespsp} since the background of these
modes from radiative tail of the $\psp$ is at the order of 0.1~pb. But
for those modes which have large branching fractions in $\psp$
decays, e.g. $b_1\pi$, the contribution of the $\psp$ tail must be
considered carefully.

\section{Summery}

In this paper, we examined the radiative correction and Monte
Carlo simulation for the measurement of $\pspp$ exclusive decays
into light hadrons in $\EE$ experiments. We draw special attention on
the interference effect between the $\pspp$ resonance and 
non-resonance continuum amplitudes. We analyzed how the
interference, particularly the destructive interference may change
the invariant mass distribution of the final state hadrons. The
discussions also lead to a possible scheme to verify the
destructive interference using only the data taken at the energy
of the $\pspp$ mass for the detectors with the energy-momentum
resolution of $(1 \sim 2)\%$. We suggest this scheme be applied on
decays such as $\omega\eta$, $\omega\eta^\prime$, $\KKSC$, and
$\rhopi$ by future BESIII experiment.

\appendix

\section{The hadronic current}

The current of virtual photon transition to the final state $f$ is
defined as~\cite{martin}
$$ {\cal H}_{\mu} \equiv \langle f |_{out} J^{em}_{\mu} |0 \rangle. $$
For reference we include its forms for some final hadronic states in the 
following. A complete list of ${\cal H}_{\mu}$ for all possible two-body 
final states can be found in Refs.~\cite{ptsai2,tosa}.

\subsection{Pseudoscalar-pseudoscalar}

For the final states $\pi^+\pi^-$ or $K^+K^-$,
 $$
 {\cal H}_{\mu} = {\cal F}_{P}(s)(p_{+}-p_{-})_{\mu},
 \label{pi}
 $$
where $p_+$ and $p_-$ are the four-momentum vectors of $\pi^+$
($K^+$) and $\pi^-$ ($K^-$) respectively, and ${\cal F}_{P}(s)$ is
the $\pi$ or $K$ form factor at the energy scale $s=(p_+ + p_-)^2$.

\subsection{Vector-pseudoscalar}

For the VP final states,
 $$
 {\cal H}_{\mu} = {\cal F}_{VP}(s)\epsilon_{\mu\alpha\beta\gamma}
 p_{V}^{\alpha}p_{P}^{\beta}e^{\gamma},
 $$
with $\epsilon_{\mu\alpha\beta\gamma}$ the completely
antisymmetric unit tensor of fourth rank, $p_{V}$ and $p_{P}$ are
the four-momentum vectors of the vector and pseudoscalar mesons
respectively, $e$ is the polarization of the vector meson, ${\cal
F}_{VP}(s)$ is the form factor at the energy scale
$s=(p_{V}+p_{P})^2$.

\subsection{$\pi^+\pi^-\pi^0$ with $\rhopi$ intermediate states}

For the 3-body final states $\pi^+\pi^-\pi^0$,
 $$
 {\cal H}_{\mu} = {\cal F}_{3\pi}(s)\epsilon_{\mu\alpha\beta\gamma}
 p_{+}^{\alpha}p_{-}^{\beta}p_{0}^{\gamma},
 $$
with $p_+$, $p_-$ and $p_0$ the four-momentum vectors of $\pi^+$,
$\pi^-$ and $\pi^0$ respectively, ${\cal F}_{3\pi}(s)$ the form
factor at the energy scale $s=(p_++p_-+p_0)^2$.

We may also include the three intermediate states $\rho^+\pi^-$,
$\rho^-\pi^+$, $\rho^0\pi^0$ and their interference into ${\cal
H}_\mu$ by multiplying a factor
\begin{widetext}
 $$
   \frac{m_\rho^2}{p_{+-}^2-m_\rho^2+i\Gamma_\rho(p_{+-}^2) s /m_\rho}
 + \frac{m_\rho^2}{p_{+0}^2-m_\rho^2+i\Gamma_\rho(p_{+0}^2) s /m_\rho}
 + \frac{m_\rho^2}{p_{-0}^2-m_\rho^2+i\Gamma_\rho(p_{-0}^2) s /m_\rho},
 $$
\end{widetext}
where $m_\rho$ and $\Gamma_\rho(s)$ are the mass and
energy-dependent width of $\rho$, $p_{+-}=p_{+}+p_-$,
$p_{+0}=p_{+}+p_0$, and $p_{-0}=p_{-}+p_0$.

\section{Solutions of $|{\cal C}|$ and $\phi$}

In this section, we discuss the possible solutions of $|{\cal C}|$ 
and $\phi$ and so the branching fraction of the resonance
decays if the experimental data are available at only two
energies, one off the resonance and the other one at the energy of
resonance mass. We begin with Eq.~\eref{int} and define
 \beq t^2 =
\left|\frac{1}{s(1-x)} + ({\cal C}+1)
\frac{3\Gamma_{ee}/\left(\alpha \sqrt{s(1-x)}\right)}
 {s(1-x)-M^2 + i M\Gamma_{t}} \right|^2~.
\label{tsqdef}
 \eeq
Here it is more convenient to consider instead of $|{\cal C}|$ and
$\phi$, the total resonance amplitude by 
 \beqns
 \rho &=&  | {\cal C} + 1 |; \\
 \psi &=&  \arg( {\cal C} + 1 )~.
 \eeqns
Here $\rho=|(\ag+\aga)/\aga|$ is the total resonance amplitude
normalized to the electromagnetic amplitude, and
$\psi=\arg((\ag+\aga)/\aga)$ is the phase of the total resonance
amplitude relative to the electromagnetic amplitude. As discussed
in Sec.~\ref{threeamplitude}, for $\pspp$, $|\aga| \ll |\ac|$, so
$\aga$ can be neglected, only the amplitudes of $\ag$ and $\ac$
are important. These two amplitude have significant interference
if their magnitude are comparable, i.e.
 $$
 |{\cal C}| \sim \frac{\alpha \Gamma_t}{3 \Gamma_{ee}}
\approx 217.
 $$
For the discussion on the interference, we only need to consider
the circumstance that $|{\cal C}| \gg 1$. Then we have
approximately $\rho \approx |{\cal C}|$ and $\psi \approx \phi$.
This means that since $|\aga| \ll |\ac|$, if $|\ag|$ is comparable
with $|\ac|$, then $\ag+\aga \approx \ag$, i.e. the total
resonance amplitude is approximately equal to the amplitude via
strong decay.

For briefness, we introduce the following notations:
 \beq
\begin{array}{c}
{\displaystyle q=\frac{1}{s(1-x)},
~~~k=\frac{3\Gamma_{ee}}{\alpha \sqrt{s(1-x)}} }, \\
{\displaystyle a=s(1-x)-M^2,~~~b=M\Gamma_{t} } ,
\end{array}
\label{eq_note}
 \eeq
then we have
 \beqn
t^2 &=&  \left| q + (|{\cal C}| e^{i \phi} +1) \cdot
 \frac{k}{a + i b} \right|^2~ \nonumber \\
    &=&  \left| q + \rho e^{i \psi}  \cdot
 \frac{k}{a + i b} \right|^2~ .
\label{t2}
 \eeqn

Eq.~\eref{t2} can be rewritten as
 \beq
 t^2 =  q^2 + 2 q R \cos
(\psi-\lambda) + R^2~~, \label{eq_tsq}
 \eeq
with
 \beqns
 R &=& {\displaystyle \frac{\rho k}{\sqrt{a^2 + b^2}} }~; \\
\tan \lambda &=&  {\displaystyle \frac{b}{a} }~.
 \eeqns

To take the radiative correction, we introduce an integral
operator defined as
 \beq
\int dG \equiv
\int \limits_{0}^{1-s_m/s} 4\pi s(1-x) \alpha^2|{\cal F}(s(1-x))|^2
{\cal P}(s(1-x))~,
 \eeq
then the radiatively corrected cross section becomes
$$ \sigma_{r.c.}(s) = \int dG t^2 \equiv T^2~.$$
Notice
$$ q R \cos (\psi -\lambda) \leq q R = \frac{qk}{\sqrt{a^2+b^2}}~, $$
and use the relation
$$ \left[ \int dG q R  \right]^2
 < \int dG q^2 \cdot \int dG \frac{k^2}{a^2+b^2}~,$$
we %immediately acquire
have
\beq
\sqrt{A^2+B^2} \cos (\psi - \Lambda) < Q K,~
\label{eq_ineq}
\eeq
where
\beqns
A = \int dG \frac{ a q k}{\sqrt{a^2+b^2}}~,~~~
B = \int dG \frac{ b q k}{\sqrt{a^2+b^2}}~, \\
K^2 = \int dG \frac{k^2}{a^2+b^2}~,~~~
Q^2 = \int dG q^2~,
\eeqns
and
$$ \tan \Lambda = \frac{B}{A}~.$$
Introduce a variable $\xi$ and let
 \beq
\cos \xi = \frac{\sqrt{A^2+B^2}}{Q K} \cdot \cos (\psi - \Lambda)~,
\label{eq_xi}
 \eeq
we obtain an expression similar to Eq.~\eref{eq_tsq}: %that is
 \beq
T^2 =  Q^2 + 2 \rho Q K \cos \xi + \rho^2 K^2~~.
\label{eq_lagtsq_1}
 \eeq
In virtue of Eq.~\eref{eq_xi}, the angle $\xi$ does not have an
apparent physical meaning but has a rather complex relation to the
angle $\psi$.

\begin{figure}[tbh]
\begin{minipage}{8.8cm}
\includegraphics[width=3.5cm,height=2.6cm]{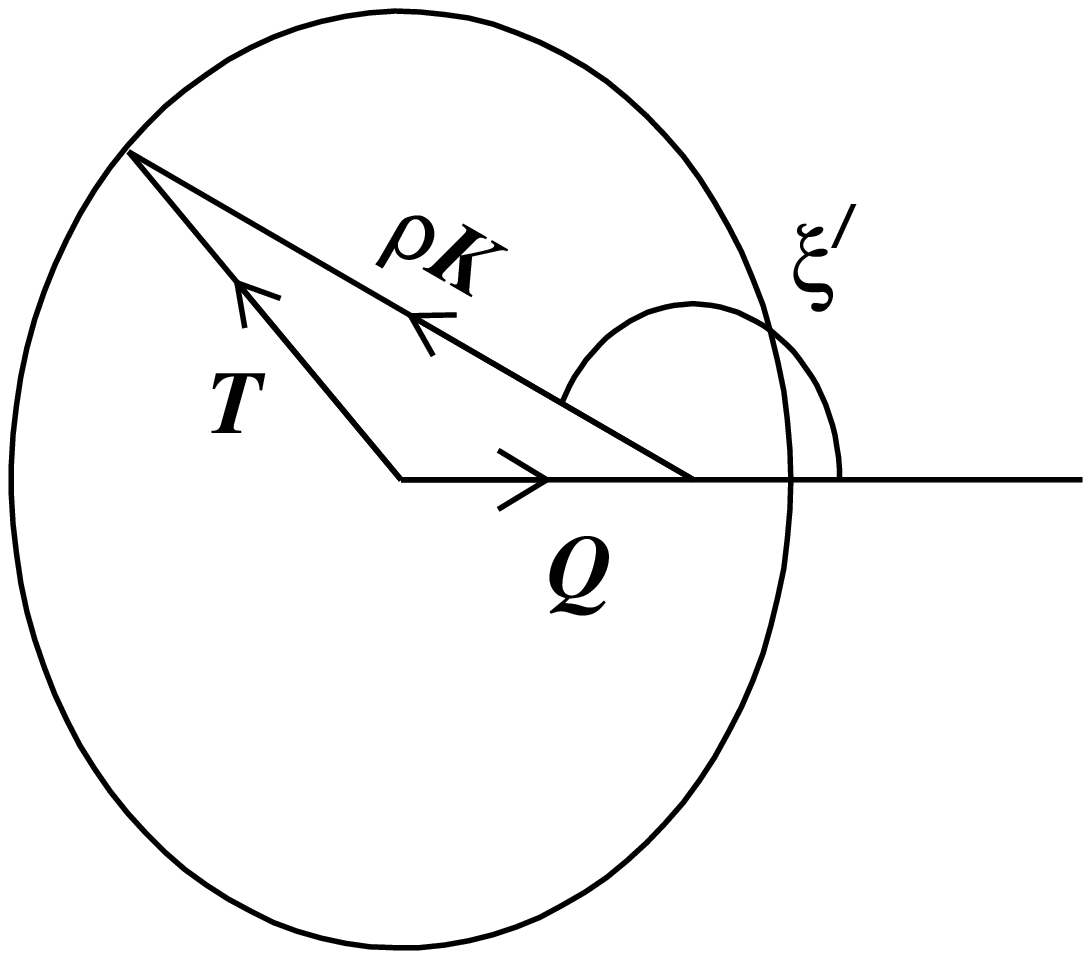}
\hskip 0.5cm
\includegraphics[width=4.2 cm,height=2.6cm]{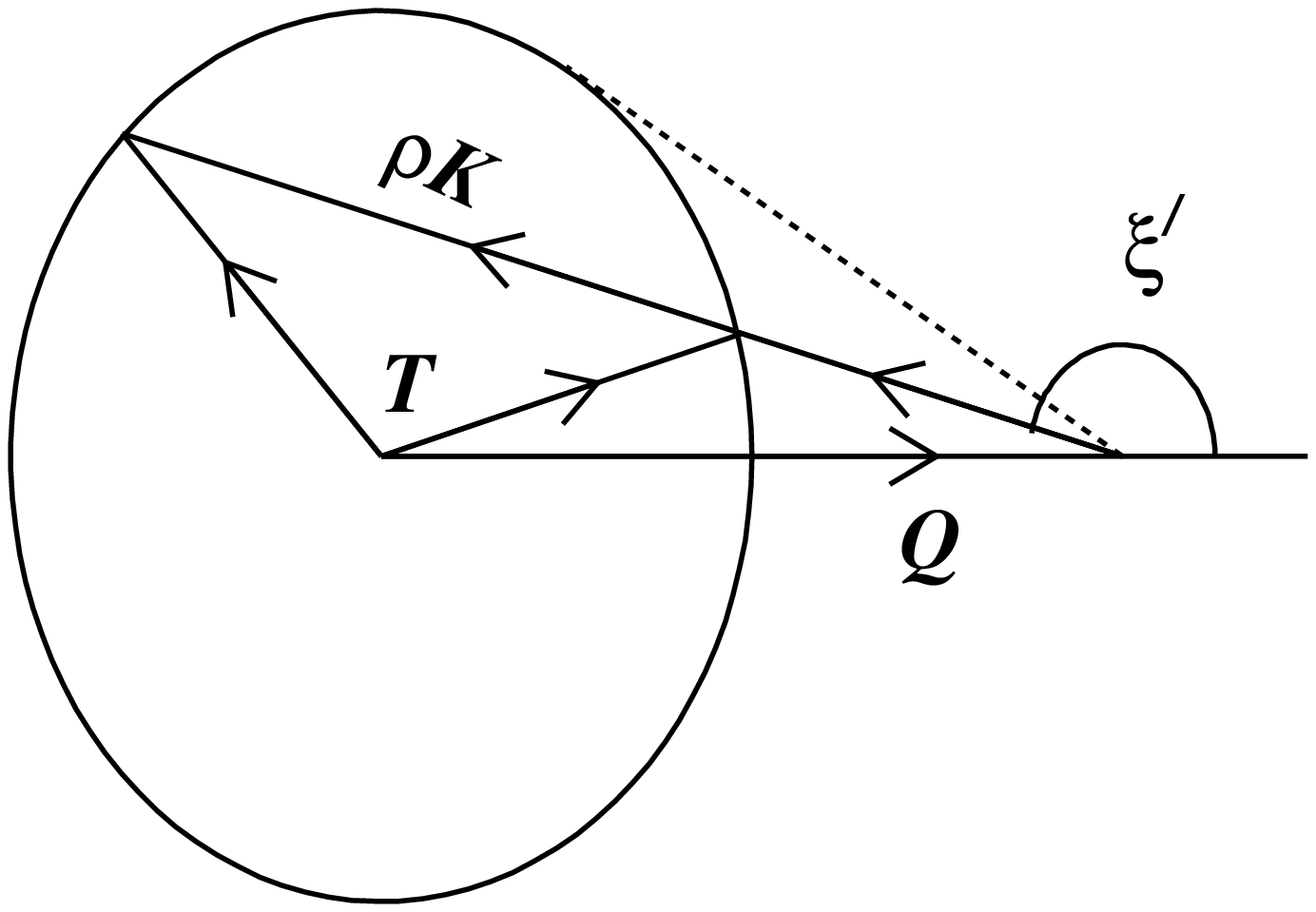}
(a) $T>Q$ \hskip 2.5cm (b) $T<Q$
\end{minipage}
\caption{\label{fig_slun} Possible solutions of $\rho$ and $\psi$
for $T>Q$ (a) and for $T<Q$ (b). Here $T,~Q,~\rho K$ denote the
total, continuum and resonance amplitudes, respectively. In the
figures $\xi^{\prime}= \psi^{\prime}-\Lambda$, where
$\psi^{\prime}= - \psi$ according to the angle definition between
two vetors. In (b) the dashed line corresponds to one solution case
for special $\xi^{\prime}$. }
\end{figure}

Introduce an angle $\zeta$ defined as $\zeta=\pi - \xi$, then
Eq.~\eref{eq_lagtsq_1} becomes,
 \beq T^2 =  Q^2 - 2 \rho Q K \cos
\zeta + \rho^2 K^2~~, \label{eq_lagtsq}
 \eeq
in which the three variables $T$, $Q$ and $\rho K$ form a triangle
and obey the law of cosines. If $T > Q$, $\zeta$ may take any
value between $-180^\circ \sim 180^\circ$, there is one and only
one solution for a given value of $\zeta$; while if $T < Q$, there
are two solutions for a value of $\zeta$, but the range of
$\zeta$ is constrained by $\sin \zeta \le T/Q$. For $\sin \zeta =
Q/T$, the two solutions coincide. In Figs.~\ref{fig_slun}(a) and
\ref{fig_slun}(b) we show schematically
the two cases for $T > Q$  and $T < Q$ respectively.

Next we consider the variation of $\epsilon(s_m)$. As discussed in
Sec.~\ref{sec_mint}, to determine the efficiency by the Monte
Carlo simulation, the resonance parameter $\rho$ and $\psi$ are
input parameters which are also the quantities to be measured. So
the data analysis is an iterative process. Here we are to prove
that if the efficiency
%is a weak-dependent function of $\rho$ and/or $\psi$,
remains stable for different solutions of $\rho$ and $\psi$, or
with only weak dependence on these two variables, the above
discussion is still valid.

In Eq.~\eref{eq_lagtsq}, $T^2$ is the total observed cross
section, which is obtained by experimentally measured quantities
as
$$ T^2 =\frac{N}{{\cal L} \cdot \epsilon }~,$$
where the efficiency $\epsilon$ is a function of $\rho$ and $\psi$.
Without losing generality, we only write its dependence on $\rho$
explicitly, i.e. $\epsilon=\epsilon(\rho)$.
%As we assume,
As shown in Sec.~\ref{sec_mint}, $\epsilon(\rho)$ remains stable
for different solutions of $\rho$ and $\psi$, and the dependence
on these variables is weak, so we take Taylor expansion of
$\epsilon$ in terms of $\rho$, and neglect the higher order terms:
 \beqns
\epsilon(\rho) &=& \epsilon(\rho_0) + {\displaystyle \left.
\frac{d \epsilon(\rho)}{d \rho } \right|_{\rho=\rho_0} (\rho-\rho_0) }+
 {\cal O} [(\rho-\rho_0)^2]  \\
          &\approx & \epsilon_0 + \eta \rho
=\epsilon_0 {\displaystyle \left(1+ \frac{\eta}{\epsilon_0}\rho\right)}~,
 \eeqns
where
 \beqns
\epsilon_0 &=& \epsilon(\rho_0) - {\displaystyle \left.
\frac{d \epsilon(\rho)}{d \rho } \right|_{\rho=\rho_0} \rho_0 }~,\\
\eta &=& {\displaystyle \left.
\frac{d \epsilon(\rho)}{d \rho } \right|_{\rho=\rho_0} }~.
 \eeqns
Here the correction term to $\epsilon_0$ is small, i.e.
 \beq
  \frac{\eta \rho}{\epsilon_0} \ll 1~.
\label{taylor1}
 \eeq
Therefore we have
%%\beqns
$$
T^2 = \frac{N}{{\cal L} \cdot {\displaystyle \epsilon_0 \cdot
\left(1+ \frac{\eta \rho}{\epsilon_0}\right) }}
    = \frac{N}{{\cal L} \cdot \epsilon_0} \cdot
{\displaystyle \left(1- \frac{\eta \rho}{\epsilon_0}\right)}
    = \sigma_t \cdot \left(1- \frac{\eta \rho}{\epsilon_0}\right),
$$
%%\eeqns
where we define
$$ \sigma_t = \frac{N}{{\cal L} \cdot \epsilon_0}~,$$
with $\sigma_t$ the total cross section calculated by the
efficiency $\epsilon_0$. $Q^2$ is the continuum cross section
which does not depend on $\rho$ and $\psi$. Substituting the above
expression of $T^2$ back into Eq.~\eref{eq_lagtsq} we have
 \beq
K^2 \rho^2 -\left( 2QK\cos\zeta-\sigma_t \frac{\eta}{\epsilon_0}
\right)
 \rho + (Q^2 - \sigma_t) = 0~.
\label{eq_rho}
 \eeq
We get
\beq
\rho=\left[(Q \cos \zeta - \delta \sigma_t) \pm
\sqrt{(Q \cos \zeta - \delta \sigma_t)^2+(\sigma_t-Q^2)}\right]/K~,
\label{eq_rhoslun}
\eeq
with
\beq
\delta = \frac{\eta}{2 K \epsilon_0}~.
\label{def_delta}
\eeq
Consider the quantity under the radical
\beqns
z&=&(Q \cos \zeta - \delta \sigma_t)^2+(\sigma_t-Q^2) \\
 &=& \delta^2 \sigma^2_t - (2 \delta Q \cos \zeta -1)\sigma_t -
 Q^2 \sin^2 \zeta~.
\label{z}
 \eeqns
In order for Eq.~\eref{eq_rhoslun} to have solutions, it requires
$z \ge 0$. For $\sigma_t > Q^2$, this is always true.
Notice that %in Eq.\eref{eq_rhoslun},
$\rho$ must be greater than 0 by definition, in this circumstance,
$\rho$ has one and only one solution for any given value of
$\zeta$, with no constraint on $\zeta$. If $\sigma_t < Q^2$, in
order to have $z>0$, we need the condition
$$\sigma_t >\frac{(2 \delta Q \cos \zeta -1) +
\sqrt{(2 \delta Q \cos \zeta -1)^2+4 \delta^2 Q^2 \sin^2 \zeta} }
{2 \delta^2}~ $$ which imposes constraint on $\zeta$. In such
case, there are two solutions for each allowed value of $\zeta$.
In Eq.~\eref{def_delta}, $\eta/\epsilon_0$ is small, so is
$\delta$. In the limit $\delta \to 0$, we find $\rho$ has two
solutions when $ Q^2 \sin^2 \zeta < \sigma_t < Q^2$. If $z=0$,
which means
$$\sigma_t =\frac{(2 \delta Q \cos \zeta -1) +
\sqrt{(2 \delta Q \cos \zeta -1)^2+4 \delta^2 Q^2 \sin^2 \zeta} }
{2 \delta^2}~, $$ the two solutions of Eq.~\eref{eq_rho} becomes
one.

%%%%%%%%%%%%%%%%%%%%%%%%%%%%%%%%%%%%%%%%%%%%%%%%%%%%%%%%%%%%

%%%%%%%%%%%%%%%%%%%%%%%%%%%%%%%%%%%%%%%%%%%%%%%%%%%%%%%%%%%%


\begin{thebibliography}{99}

\bibitem{rosner}J.L.~Rosner,
  \Journal\PRD{64}{094002}{2001}.
\bibitem{wym7}P.~Wang, X.H.~Mo and C.Z.~Yuan,
  \Journal\PRD{70}{077505}{2004}.
\bibitem{wym1}P.~Wang, C.Z.~Yuan and X.H.~Mo,
  \Journal\PRD{70}{114014}{2004}.
\bibitem{cleolp05}CLEO Collaboration, G.S.~Adam {\em et al.},
  hep-ex/0509011.
\bibitem{besvp}BES Collaboration, M.~Ablikim {\em et al.},
  \Journal\PRD{70}{112007}{2004}.
\bibitem{besrp}BES Collaboration, M.~Ablikim {\em et al.},
  \Journal\PRD{72}{072007}{2005}.
\bibitem{wym3}P.~Wang, C.Z.~Yuan and X.H.~Mo,
  \Journal\PLB{574}{41}{2004}.
\bibitem{wymz}P.~Wang, C.Z.~Yuan, X.H.~Mo and D.H.~Zhang,
  \Journal\PLB{593}{89}{2004}.
\bibitem{rudaz}S.~Rudaz, \Journal\PRD{14}{298}{1976}.
\bibitem{phyreport}L.~K$\ddot{\hbox{o}}$pke and N.~Wermes,
           \Journal\PRP{174}{67}{1989}.
\bibitem{haber}H.E.~Haber and J.~Perrier,
  \Journal\PRD{32}{2961}{1985}.
\bibitem{bes2pspkskl}BES Collaboration, J.Z.~Bai {\em et al.},
       \Journal\PRL{92}{052001}{2004}.
\bibitem{cleoformf}CLEO Collaboration, G.S.~Adam {\em et al.},
  hep-ex/0510005.
\bibitem{Zline}F.A.~Berends {\em et al.}, in
Proceedings of the Workshop on Z Physics at LEP, v.1, (1989) page
89, edited by G.~Altarelli, R.~Kleiss and C.~Verzegnassi.
\bibitem{pdg}Particle Data Group, S.~Eidelman {\em
et al.}, \Journal\PLB{592}{1}{2004}.
\bibitem{wym4}P.~Wang, C.Z.~Yuan and X.H.~Mo,
  \Journal\PRD{69}{057502}{2004}.
\bibitem{wym2}P.~Wang, C.Z.~Yuan and X.H.~Mo,
  \Journal\PLB{567}{73}{2003}
\bibitem{ptsai}Y.S.~Tsai, SLAC-PUB-3129 (1983).
\bibitem{rad}E.A.~Kuraev and V.S.~Fadin, Yad. Fiz. {\bf 41}
       (1985) 733 [Sov. J. Nucl. Phys. {\bf 41} (1985) 466];
    G.~Altarelli and G.~Martinelli, CERN {\bf 86-02} (1986) 47;
    O.~Nicrosini and L.~Trentadue, \Journal\PLB{196}{551}{1987};
    F.A.~Berends, G.~Burgers and W.L.~Neerven, \Journal\NPB
    {297}{429}{1988}; {\it ibid.} {\bf 304} (1988) 921.
\bibitem{Greco} A.~Bramon and M.~Greco, in {\it The Second
    DA$\Phi$NE Physics Handbook}, edited by I.Maiani,
    G.~Pancheri and N.~Paver, Vol.2, p451, 1995.
%\bibitem{vacuum}F.A.~Berends and G.L.~Komen,
%    \Journal\PLB{63}{432}{1976}.
\bibitem{hard1}C.M.~Carloni, \Journal\PLB{520}{16}{2001}.
\bibitem{hard2} A.B.~Arbuzov {\em et al.}, hep-ph/0504233;
   A.B.~Arbuzov {\em et al.}, \Journal\JHEP{10}{006}{1997}.
\bibitem{ISR}H.~Czy\.{z} and J.H.~K\"{u}hn,
    \Journal\EPJC{18}{497}{2001}.
\bibitem{jadach}J.~Jadach, B.F.L.~Ward and Z.~Was,
        \Journal\PRD{63}{113009}{2001}.
\bibitem{cleopsp}CLEO Collaboration, N.E.~Adam {\em et al.},
   \Journal\PRL{94}{012005}{2005}.
\bibitem{bespsp}BES Collaboration, M.~Ablikim {\em et al.},
   \Journal\PLB{614}{37}{2005}; \\
   BES Collaboration, M.~Ablikim {\em et al.},
   \Journal\PLB{619}{247}{2005};
\bibitem{martin}G.~Bonneau and F.~Martin, \Journal\NPB{27}{381}{1971}.
\bibitem{ptsai2}Y.S.~Tsai, \Journal\PRD{12}{3533}{1975}.
\bibitem{tosa}Y.~Tosa, Nagoya University preprint DPNU-34(1976).
\end{thebibliography}
\end{document}